\documentclass[%
aip,
 amsmath,amssymb,
floatfix
]{revtex4-1}

\usepackage{graphicx}% Include figure files
\usepackage{dcolumn}% Align table columns on decimal point
\usepackage{bm}% bold math
\usepackage[utf8]{inputenc}
\usepackage[T1]{fontenc}
\usepackage{mathptmx}

\usepackage{xcolor}	% DELETE, IF IT IS NOT NECESSARY!
\usepackage{subfig}
\usepackage{csquotes}
\usepackage{lmodern}
\usepackage[T1]{fontenc}
\usepackage[colorlinks=true,linkcolor=blue,allcolors=blue]{hyperref}%

\begin{document}

%\preprint{AIP/123-QED}

\title{First close-coupling study of the excitation of a large cyclic molecule: collision of \textbf{c}-C\textsubscript{5}H\textsubscript{6} with He}
% Force line breaks with \\

\author{S\'{a}ndor Demes}
  \email{sandor.demes@univ-rennes.fr}
\author{Cheikh T. Bop}%
\affiliation{Univ Rennes, CNRS, IPR (Institut de Physique de Rennes) - UMR 6251, F-35000 Rennes, France}%
\author{Malek Ben Khalifa}
%% \email{claire.rist@univ-grenoble-alpes.fr}
\affiliation{KU Leuven, Department of Chemistry, B-3001 Leuven, Belgium}%
\author{Fran\c{c}ois Lique}%
\affiliation{Univ Rennes, CNRS, IPR (Institut de Physique de Rennes) - UMR 6251, F-35000 Rennes, France}%

{Published in {\it Phys. Chem. Chem. Phys} at \url{https://doi.org/10.1039/D4CP01380H}}
\date{24 May 2024}

\begin{abstract}
Recent astronomical observations revealed an increasing molecular complexity in the interstellar medium through the detection of a series of large cyclic carbon species. To correctly interpret these detections, a complex analysis is necessary that takes into account the non-local thermodynamic equilibrium (non-LTE) conditions of the emitting media ({\it e.g.} when energy level populations deviate from a Boltzman distribution). This requires proper state-to-state collisional data for the excitation and de-excitation processes of the molecular levels. Cyclopentadiene ($c$-C$_5$H$_6$), which was recently detected in cold interstellar clouds, is extensively studied in many aspects due to its large importance for chemistry in general. At the same time, there are no collisional data available for this species, which are necessary for a more precise interpretation of the corresponding detections. In this work, we first provide an accurate 3D rigid-rotor interaction potential for the [$c$-C$_5$H$_6 +$He] complex from high-level of {\it ab initio} theories, %It is highly anisotropic, with several local minima and a global well depth of about $-90$ cm$^{-1}$. Once the potential was fitted over an analytical angular expansion,it
which has been used to study their inelastic collision by the exact close coupling quantum scattering method. To the best of our knowledge, this is the first study where this method is systematically applied to treat the dynamics of molecular collisions involving more than ten atoms. We also analyse the collisional propensity rules and the differences in contrast to calculations, where the approximate coupled states scattering methods is used.
\end{abstract}

\maketitle

\section{\label{sec:Intro} Introduction}

Cyclopentadiene ($c$-C$_5$H$_6$) and its functionalized derivatives are widely used in synthetic organic and organometallic chemistry.\cite{dalkilic2015} It is a very reactive diene, which can easily form dimer ring structures by Diels-Alder reactions. Besides that, its radical C$_5$H$_5$ formed by H-loss is expected to be a key precursor to polycyclic aromatic hydrocarbons (PAHs).\cite{wang2006}
%He et al. (2021) \cite{he2021} showed that C$_5$H$_6$ can be effectively formed in the gas phase from 1,2-butadiene (CH$_2$CCHCH$_3$) in such low-temperature environments, which are specific to cold interstellar molecular clouds.
As shown by Chiar \textit{et al.} (2013),\cite{chiar2013} large cyclic species and other complex organic molecules (COMs) are expected to be widespread in the interstellar medium (ISM), building up simpler five- and six-membered carbon rings but also complex PAHs. These cycles are also known to be the building blocks of organic chemistry on earth. Despite their expected prevalence, the astronomical detection in space has proven to be a difficult task for a while, due to their relatively small dipole moment and dense rotational spectra. However, the widespread presence of carbon cycles in the ISM are proven now by a series of recent observations in the mid-infrared spectral range. The first discovery of benzonitrile ($c$-C$_6$H$_5$CN), one of the simplest detectable aromatic species, has been reported by McGuire \textit{et al.} (2018).\cite{mcguire2018} The detection and importance of the five-membered $c$-C$_5$H$_6$ in the interstellar medium also have been the subject of several recent studies, shedding light on its role in interstellar chemistry and its potential implications for our understanding of complex organic molecules in space. First, McCarthy \textit{et al.} (2020) \cite{mccarthy2020} confirmed the presence of the most stable isomer of $c$-C$_5$H$_5$CN, 1-cyano-1,3-cyclopentadiene, which is a highly polar cyano derivative of $c$-C$_5$H$_6$. Another isomer, the 2-cyano-1,3-cyclopentadiene has been detected soon by the same group.\cite{kelvin2021} The first discovery of cyclopentadiene itself, along with other cyclic hydrocarbons such as $c$-C$_3$HCCH and $c$-C$_9$H$_8$, has been reported at the same time by Cernicharo \textit{et al.} (2021).\cite{cernicharo2021} In a very recent observational study by Ag\'{u}ndez \textit{et al.} (2023),\cite{agundez2023} an evidence was found once again that cyclic hydrocarbons, including $c$-C$_5$H$_6$, $c$-C$_6$H$_5$CN and $c$-C$_5$H$_5$CN, are widespread in cold interstellar clouds. These discoveries contribute to our understanding of the chemical networks of carbon rings in interstellar environments and highlight the importance of $c$-C$_5$H$_6$ in these processes. In general, the detections of cyclopentadiene and its derivatives in the ISM have been an important milestone in observational astronomy by confirming the widespread existence of complex, cyclic hydrocarbons in space. In addition, they also provide insights into the chemical pathways leading to the formation of PAHs and COMs in molecular clouds and star-forming regions. It is worth to mention that all these astronomical surveys targeted cold clouds of the ISM, most notably the prestellar core of the Taurus Molecular Cloud (TMC-1), where the usual temperature is around $10$~K. Further research is obviously needed however to uncover additional details about the abundance, spatial distribution, and reactivity of $c$-C$_5$H$_6$ and other interstellar carbon cycles in various regions of the ISM.

\begin{figure*}[ht]
\centering
  \includegraphics[width=1\linewidth]{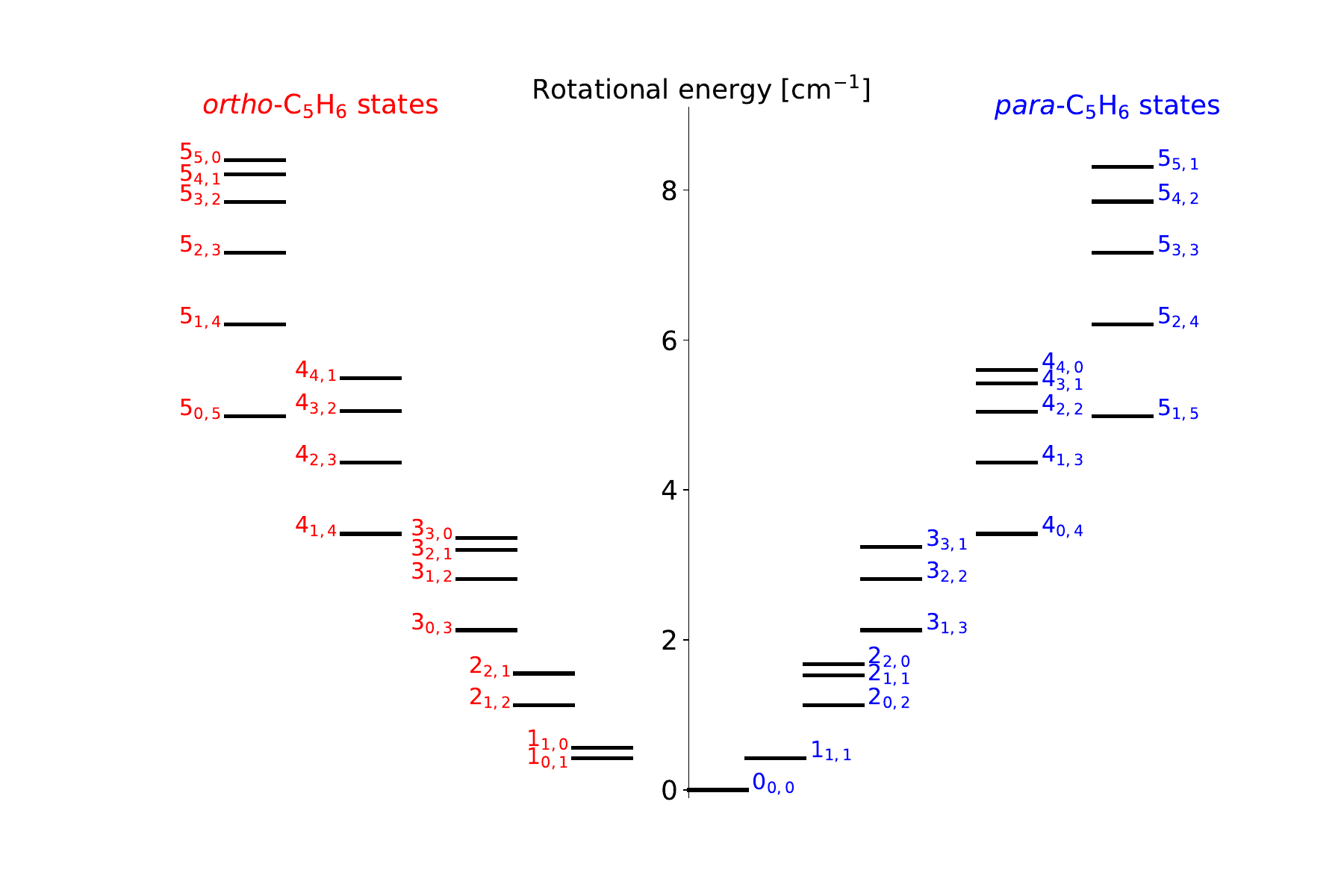}
  \caption{All rotational levels of the {\it ortho}- and {\it para}-C$_5$H$_6$ with $j \leq 5$.}
  \label{fig:rotlev}
\end{figure*}

The microwave spectrum of cyclopentadiene is very well studied.\cite{laurie1956,Damiani_1976,bogey1988} For example, in a rotational spectroscopy measurement,\cite{laurie1956} it was shown that it has a $0.416$~D dipole moment  along the $b$ rotational axis. Due to its $C_{2v}$ symmetry, it is an asymmetric rotor with rotational levels defined as $j_{k_a,k_c}$, where the sum of rotational projection quantum numbers $k_a + k_c$ being odd or even defines the {\it ortho}($o$) and {\it para}($p$) nuclear spin configurations, respectively. For example, the first  {\it ortho}-level $j_{k_a,k_c} =  1_{0,1}$ is only about 0.42 cm$^{-1}$ higher than the ground-state of $p$-C$_5$H$_6$ ($j_{k_a,k_c} = 0_{0,0}$). Since its rotational constants are low ($A = 0.281$~cm$^{-1}$, $B = 0.274$ cm$^{-1}$ and $C = 0.142$ cm$^{-1}$),\cite{bogey1988} the levels in cyclopentadiene are densely packed (see Fig~\ref{fig:rotlev} for details). In general, most of the {\it o}/{\it p}-C$_5$H$_6$ states are very close in energy due to the nearly degenerated $A$ and $B$ rotational constants, but {\it ortho}-to-{\it para} transitions are forbidden due to the spin-selectivity rules.

For proper interpretation of astronomical observations, accurate spectroscopic information alone is usually not sufficient. For example, in interstellar environments where the density is not high enough to maintain local thermodynamic equilibrium, the radiative and collisional processes compete to populate the molecular levels. Proper modelling of the molecular spectra then requires knowledge of the state-to-state rate coefficients for the collisional excitation and de-excitation processes. At the same time, only a very few theoretical works have been devoted to studying the collisional properties of nonlinear interstellar species containing more than 4 atoms, induced by atomic or molecular projectiles. Among them, mostly symmetric top species have been studied by the exact close coupling (CC) quantum scattering method, such as the acetonitrile (CH$_3$CN) and methyl isocyanide (CH$_3$NC) \cite{BenKhalifa_2022} or propyne (CH$_3$CCH) \cite{BenKhalifa_2024} in collision with He, as well as methanol (CH$_3$OH) due to H$_2$ impact.\cite{Dagdigian_2023} In the case of asymmetric top molecules, one should mention the collisional excitation of cyclic carbene ($c$-C$_3$H$_2$) by H$_2$ studied by the CC method,\cite{Khalifa2019} nearly simultaneously with propylene oxide (CH$_3$CHCH$_2$O),\cite{Faure_2019} which has been investigated in collision with He, using the approximate coupled states (CS) method (for more details regarding the differences between the CC and CS methods, see Section~\ref{sec:ScattMeth}). The first collisional data for a large cyclic species has been recently reported by Mandal \textit{et al.} (2022).\cite{mandal2022} The authors studied the state-to-state rotational excitation and quenching of benzene (C$_6$H$_6$) by He impact, using a mixed quantum/classical theory (MQCT), covering a large range of levels ($j\leq60$) and kinetic energies ($E_c \leq 10^4$ cm$^{-1}$). The potential energy surface (PES) of the weakly bound ``benzonitrile + helium'' complex has been studied very recently in the rigid rotor approximation.\cite{Derbali_2023} Nearly simultaneously, collision-induced rotational excitation cross sections and rate coefficients have also been reported for the aromatic benzonitrile ($c$-C$_6$H$_5$CN) due to He impact,\cite{Khalifa_2023} where the authors used the CS approximation. To the best of our knowledge, there are no collisional data published at the moment for any five-membered cyclic molecules, and no systematic calculations have been carried out using the exact CC theory for any larger cyclic species in collisions with atoms or molecules.
%\textcolor{orange}{However, as it has been shown by Bop \textit{et al.}\cite{bop2022} for example, appropriate collisional data are crucial for larger species as well, to provide a proper analysis of their observational spectra under non-LTE conditions.}

The aim of the present work is to provide the first accurate collisional cross sections for the rotational excitation of cyclopentadiene induced by He impact. Helium is known to be a suitable template for H$_2$ (with a proper scaling factor) for collisional excitation of heavy molecular species.\cite{Roueff2013} The calculations are carried out by means of the close coupling method based on a new, accurate 3-dimensional PES. The paper is organized as follows. In Section~\ref{sec:PESdet} the \textit{ab initio} calculations and the potential energy surface details are discussed, Section~\ref{sec:ScattMeth} describes the scattering calculation methodologies. In Section~\ref{sec:ResDis}, we report the collisional cross sections and discuss their propensities, while our conclusions and final remarks are presented in Section~\ref{sec:Concl}.

\section{\label{sec:PESdet} Potential energy surface}

\subsection{\label{sec:ElStructCalc} Ab initio calculations}
\begin{figure}[h]
\centering
  \includegraphics[width=0.5\linewidth]{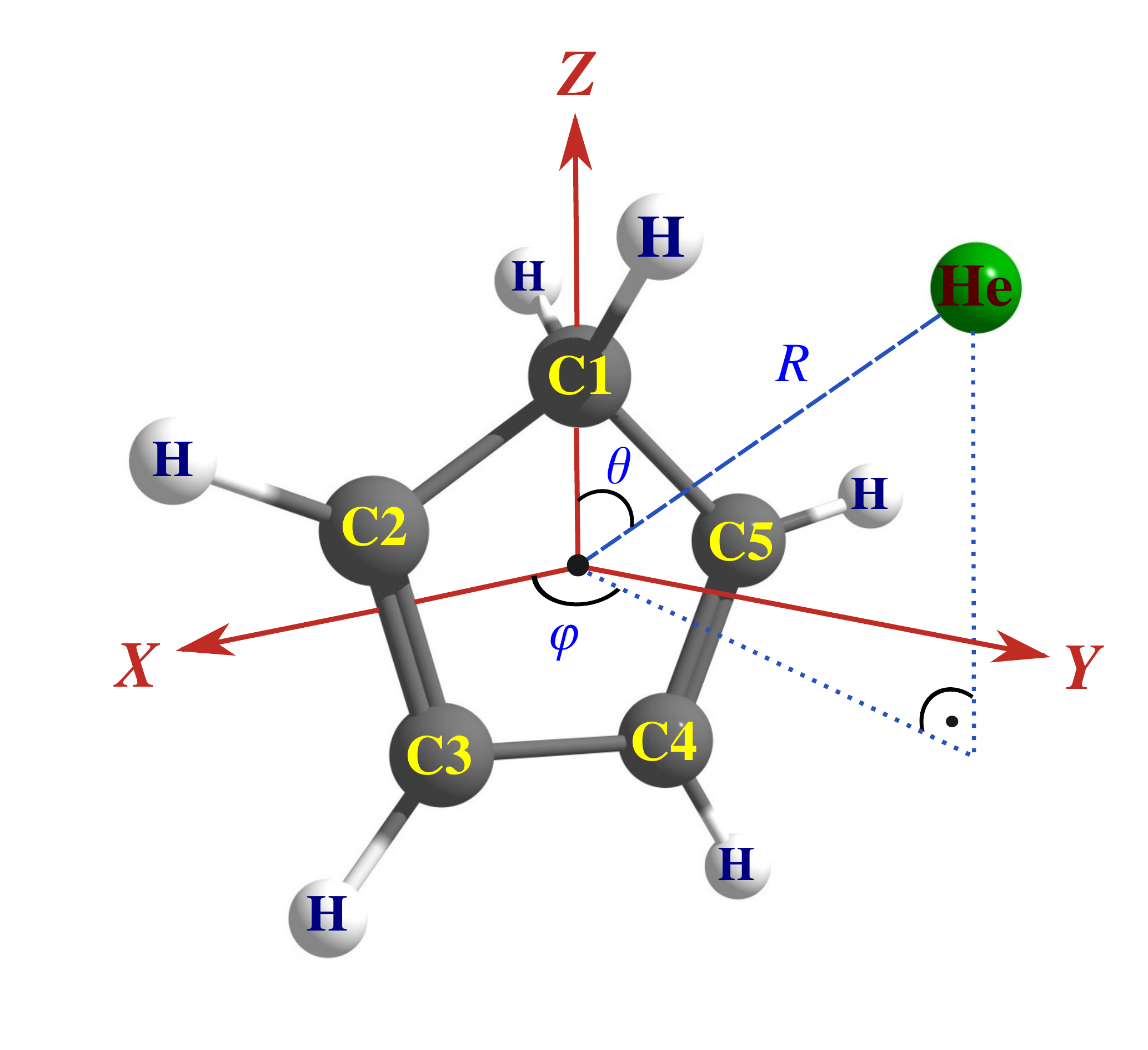}
  \caption{The Jacobi coordinate system used for the [$c$-C$_5$H$_6$ + He] complex.}
  \label{fig:coordsys}
\end{figure}

In order to treat inelastic collisions through a loosely bound intermediate complex, the knowledge of only the ground-state PES is usually adequate. We constructed an accurate adiabatic interaction potential to study the rotational excitation of $c$-C$_5$H$_6$ by He. Due to the low collision energies considered, we used a fixed intramolecular geometry for cyclopentadiene ({\it i.e.} rigid-rotor approximation). As it was shown earlier on the example of H$_2$O \cite{GarciaVasquez_2023} and HCN,\cite{DenisAlpizar_2013} the explicit consideration of the vibrational modes in the PES has no significant impact on the pure rotational transitions, even at collision energies above the first vibrational excitation threshold (note that the lowest anharmonic vibrational frequency of $c$-C$_5$H$_6$ is about $520$~cm$^{-1}$ \cite{Alparone_2006}). To define the collisional complex, a 3D Jacobi coordinate system $(R, \theta, \phi)$ has been chosen (see Fig.~\ref{fig:coordsys}). The center of the coordinate system is in the center-of-mass of the cyclopentadiene (molecular-frame representation), while the $z$-direction points towards the $a$ inertial axis and the carbon backbone of the molecule is located in the $xz$ plane. The $R$ radial parameter defines the distance between the center-of-mass of C$_5$H$_6$ and the He atom, while $\theta$ and $\phi$ refer to its relative orientation. Similar notations and coordinate system have been used  for the PES of the [$c$-C$_6$H$_5$CN + He] complex.\cite{Derbali_2023,Khalifa_2023} The equilibrium geometry of $c$-C$_5$H$_6$ have been calculated by full geometry optimization using the explicitly correlated coupled cluster theory with single and double excitation and triple corrections (CCSD(T)-F12) \cite{Adler_2007} using the \texttt{MOLPRO} quantum chemistry software package.\cite{MOLPRO_brief,Werner2012} In all calculations performed by the CCSD(T)-F12 method, the recommended augmented, correlation-consistent valence triple-$\zeta$ (aug-cc-pVTZ) basis set has been used along with standard density fitting (DF) and resolution of the identity (RI) basis terms.\cite{Dunning_1989,Weigend_2002} We found the following structural parameters: $r(\mathrm{C}_1-\mathrm{H}) = 2.0704$ $a_0$; $r(\mathrm{C}_2-\mathrm{H}) = 2.0412$ $a_0$; $r(\mathrm{C}_3-\mathrm{H}) = 2.0425$ $a_0$; $r(\mathrm{C}_1-\mathrm{C}_2) = 2.8406$ $a_0$; $r(\mathrm{C}_2=\mathrm{C}_3) = 2.55$ $a_0$; $r(\mathrm{C}_3-\mathrm{C}_4) = 2.7768$ $a_0$;
and $\angle(\mathrm{H}-\mathrm{C}_1-\mathrm{H}) = 106.64^\circ$; $\angle{(\mathrm{C}_2-\mathrm{C}_1-\mathrm{C}_5)} = 103.08^\circ$; $\angle{(\mathrm{C}_2=\mathrm{C}_3-\mathrm{C}_4)} = 109.14^\circ$; $\angle{(\mathrm{C}_1-\mathrm{C}_2=\mathrm{C}_3)} = 109.32^\circ$; $\angle{(\mathrm{C}_2=\mathrm{C}_3-\mathrm{H})} = 126.12^\circ$; $\angle{(\mathrm{C}_3=\mathrm{C}_2-\mathrm{H})} = 126.6^\circ$. These data agree within an uncertainty factor of less than $1\%$ compared to those reported by other authors for $c$-C$_5$H$_6$.\cite{Alparone_2006, Bomble_2004}
Throughout the paper, we use atomic units ($a_0$) for distances (1 $a_0 =$ 1 bohr $\approx 0.529177$ \AA), and wavenumbers (cm$^{-1}$) for energies (1~cm$^{-1} \approx 1/219474.624$ hartree).

We calculated a total of 12 540 single-point \textit{ab initio} energies for the [$c$-C$_5$H$_6 -$He] complex using the CCSD(T)-F12b/aug-cc-pVTZ level of theory. As it was shown earlier,\cite{Derbali_2023,Khalifa_2023,Demes_2020} this {\it ab initio} model is well-suited for the calculation of accurate interaction potentials. The energies have been corrected to basis set superposition error using the counterpoise procedure of Boys \& Bernardi.\cite{Boys1970} The geometries in the Jacobi coordinate system have been defined on a uniform angular grid over $0^{\circ}\leq \theta \leq180^{\circ}$ and $0^{\circ}\leq \phi \le360^{\circ}$, with a step size of $10^{\circ}$ (note that due to the $C_{2v}$ symmetry, the range of azimuthal angles can be reduced to $0^{\circ}\leq \phi \le90^{\circ}$ in the  {\it ab initio} calculations). The radial distances have spanned the interval from $R=5.0$~$a_0$ up to $R=26$~$a_0$ with a variable step size $\Delta R$, in particular $\Delta R=0.25$~$a_0$ below $15$~$a_0$, $\Delta R=0.5$~$a_0$ between $15$ and $22$~$a_0$, and $\Delta R=1.0$~$a_0$ in the long range. The collisional system has a shallow well of about $-90$ cm$^{-1}$ around $R \simeq 6.0$~$a_0$.

\subsection{\label{sec:FitPES} Fitting the PES}

For the implementation of the PES in scattering codes, a proper angular expansion is usually required over spherical harmonics functions. In the ``body-fixed'' representation for the [$c$-C$_5$H$_6 -$He] complex, the intermolecular potential can be expressed as a product of radial and angular functions as follows:

\begin{equation}
V(R, \theta, \phi) = \sum_{l}^{l_\mathrm{max}} \sum_{m}^{l} V_{lm}(R)\frac{Y_l^m(\theta, \phi) + (-1)^m  Y_l^{-m}(\theta, \phi)}{1+ \delta_{m,0}} ,
\label{eq:vrtp}
\end{equation}
where $V_{lm}(R)$ are the radial coefficients and $Y_l^m(\theta, \phi)$  are the normalized spherical harmonics, $\delta_{m,0}$ is the Kronecker $\delta$-symbol, and $l, m$ are non-negative integers, which refer to the spherical harmonics degree and order, respectively. Note that $m \leq l$ and due to the $C_{2v}$ symmetry of $c$-C$_5$H$_6$, $m$ is required to be a multiple of 2.

For all intermolecular distances $R$, the PES was fitted over an angular expansion according to Eq.~(\ref{eq:vrtp}), using a standard linear least-squares procedure. We selected a maximum order that includes anisotropies up to $l = 16$ and $m = 16$. Due to the symmetry of the target, this resulted in a total of 81 $V_{lm} (R)$ functions. Both the root mean square percentage error (RMSPE) and the mean absolute error (MAE) of the fit were found to be less then $1\%$ along the interaction potential, apart from very short distances ($R < 5.5$~$a_0$), where the PES is usually highly repulsive ($V > 10000$ cm$^{-1}$ on the average). In particular, we found the following RMSPE/MAE errors for some specific distances (MAE are given in parentheses): $R = 5.0$~$a_0$: $3.28 \%$ $(1.92 \%)$, $R = 6.0$~$a_0$: $0.52 \%$ $(0.33 \%)$, $R = 10.0$~$a_0$: $< 0.01 \%$ $(< 0.01 \%)$ and $R = 26.0$~$a_0$: $0.86 \%$ $(0.63 \%)$. At the global minimum position, the accuracy of the fit is better than $0.1$ cm$^{-1}$. To cover all radial distances required for scattering calculations, a cubic spline radial interpolation was employed over the $V_{lm} (R)$ functions in the range of $R=5-26$~$a_0$, which was smoothly connected to standard exponential and power-laws extrapolations towards the short- and long-range distances, respectively (see Ref.\cite{Valiron_2008} for details).

\begin{figure}
\includegraphics[width=0.7\linewidth]{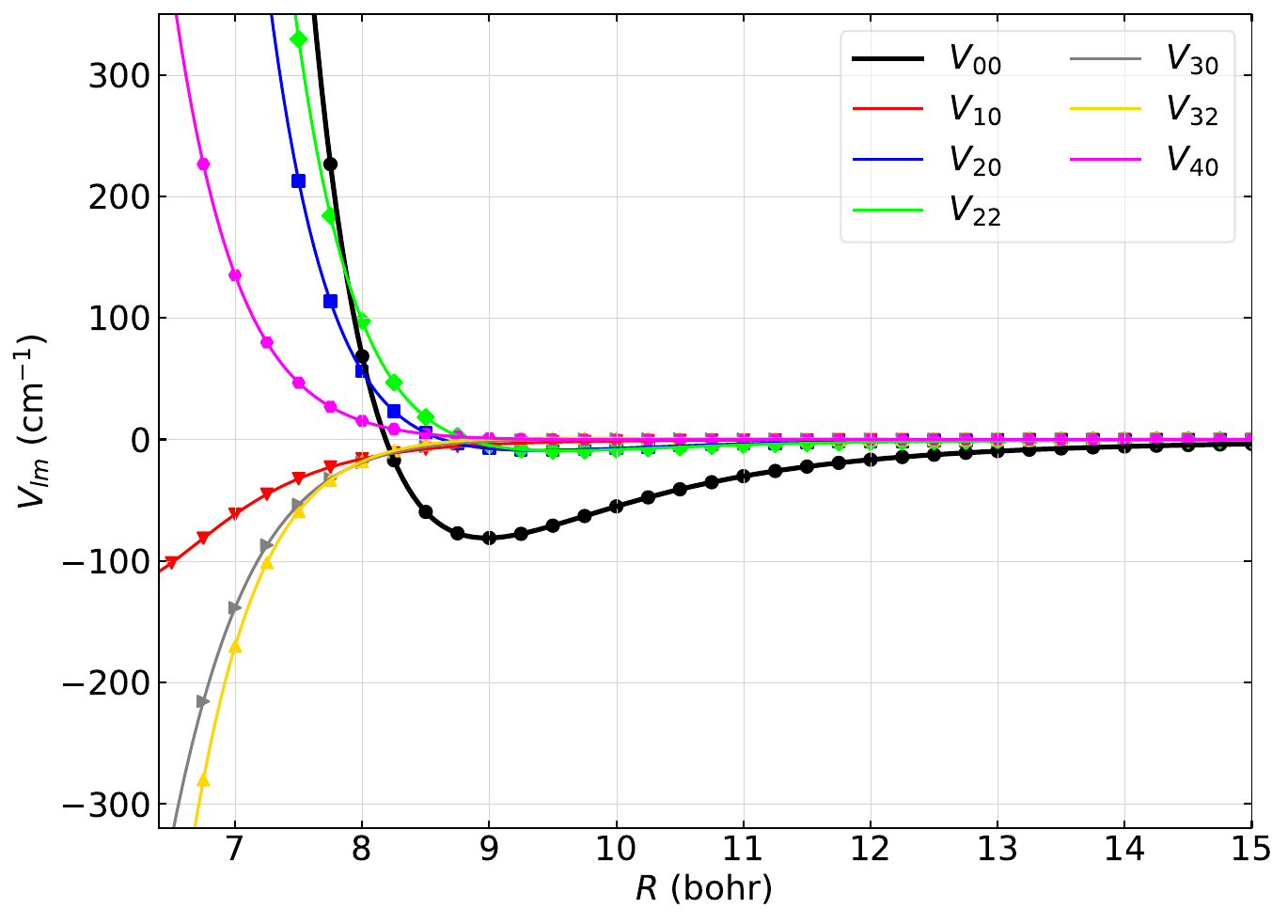}
\caption{Variation of the first 7 radial functions $V_{lm} (R)$ which were used for the analytical representation of the $c$-C$_5$H$_6$ + He  PES.}
\label{fig:vlamda}
\end{figure}

In Fig. \ref{fig:vlamda} we show the variation of some of the low-rank radial functions $V_{lm} (R)$ as a function of distance. Since the analytical fit is performed for all distances independently, it is important to examine the general $R$-dependence to benchmark its quality. Most of the terms of the potential (apart from the isotropic $V_{00}$) have either predominantly repulsive ({\it e.g. }$V_{20}, V_{22}, V_{40}$), or mostly attractive nature ({\it e.g. }$V_{30}$ and $V_{32}$), especially at short distances.

\subsection{\label{sec:StrPES} Features and structure of the PES}

\begin{figure*}
\includegraphics[width=0.49\linewidth]{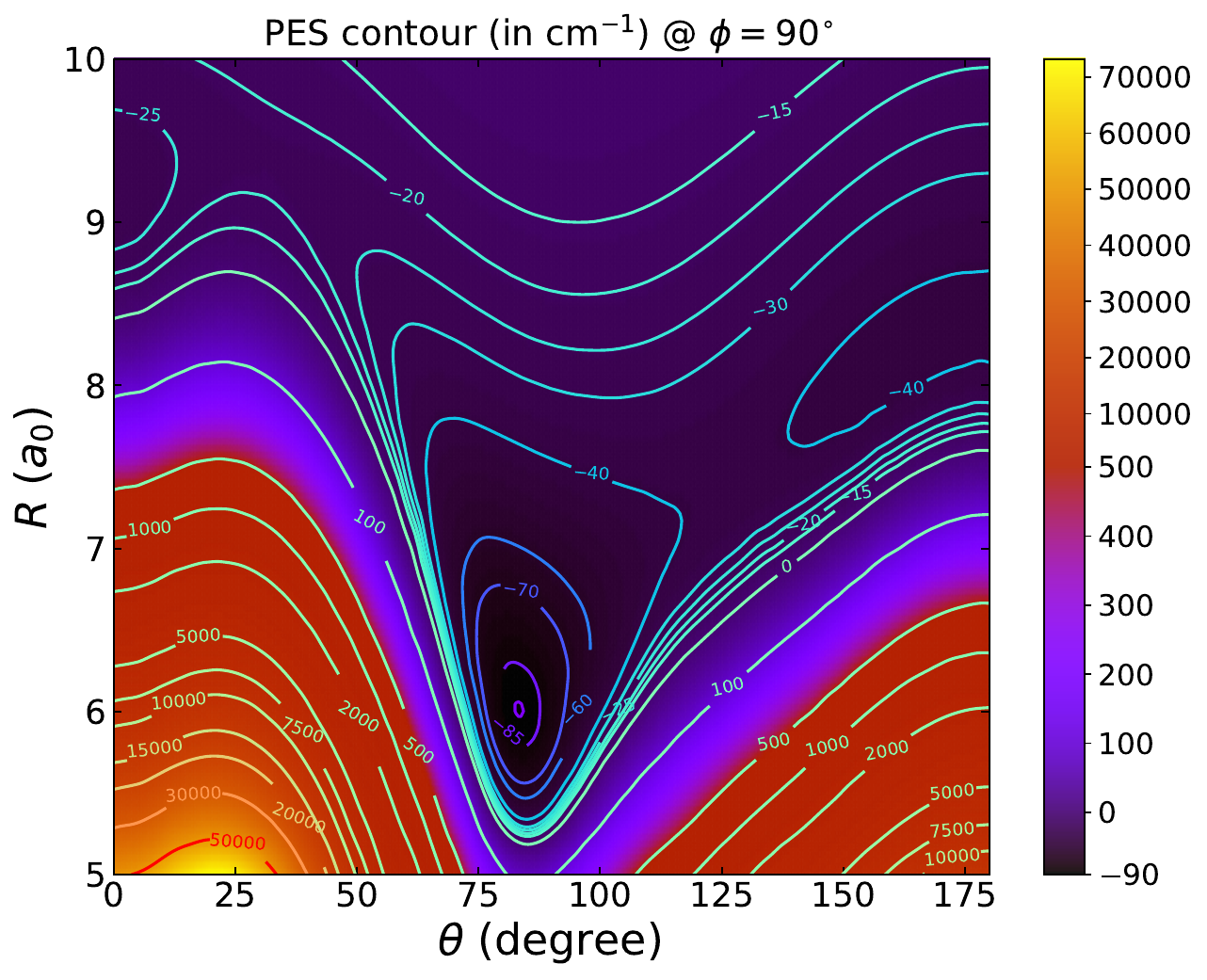}
\includegraphics[width=0.49\linewidth]{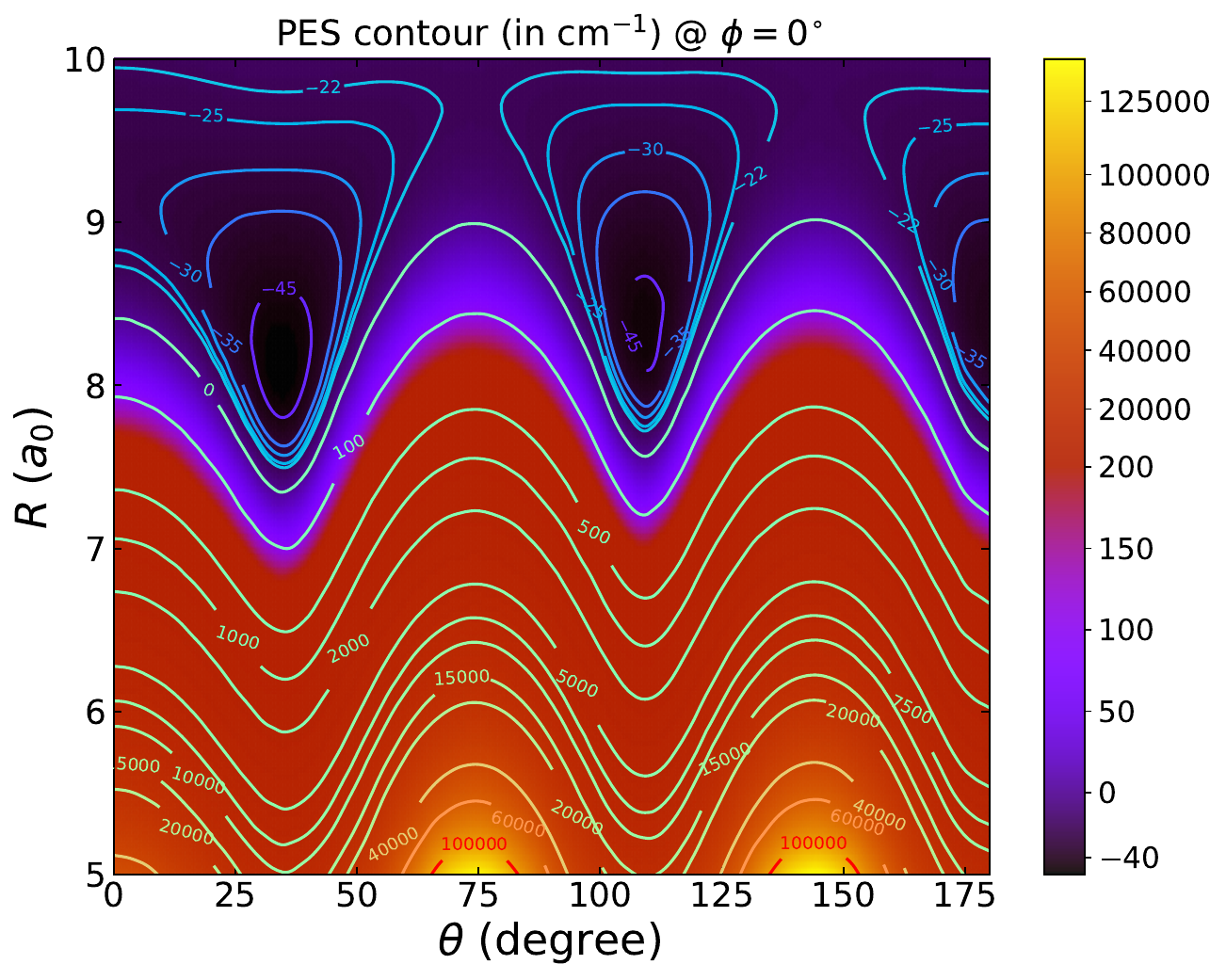}
\caption{Contour plots representing $R$ $vs.$ $\theta$ dependence of the $c$-C$_5$H$_6 + $ He PES at fixed azimuthal angles $\phi=90^{\circ}$ (left) and $\phi=0^{\circ}$ (right).}
\label{fig:PEScontRT}
\end{figure*}

For a comprehensive validation of the fit and for an illustration of the features and anisotropies of the PES, we examine some contour plots extracted from the final analytical interaction potential. In Fig.~\ref{fig:PEScontRT}, its radial dependence is shown with respect to the spherical angle $\theta$ at fixed azimuthal angles. The left subplot introduce a 2D-cut from the PES which highlights the global well located around $R \simeq 6.0$~$a_0$, $\theta \simeq 83.2^{\circ}$ and $\phi = 90^{\circ}$. As one can see, the radial dependence is very strong and characterized with extremely large gradients in energy. Apart from the global minimum, two local minima are present (at $[R \simeq 8.0$~$a_0; \theta \simeq 165^{\circ}]$ and $[R \simeq 9.5$~$a_0; \theta = 0^{\circ}]$). Changing the azimuthal angle by $\pi/2$ leads to a much different anisotropy picture (see the right subplot of Fig.~\ref{fig:PEScontRT} at $\phi = 0^{\circ}$), where the PES is more isotropic with respect to both $R$ and $\theta$, but its gradients are still very large. We can see three pronounced local minima at $R \simeq 8.5$~$a_0$, all with a depth of about $-50$ cm$^{-1}$. The shape and width of these minima are very similar to those of the global minimum. The radial behaviour of the PES is rather symmetric, and one can observe much larger anisotropies with respect to rotation for $\theta$. For example, at distances $\sim8.5$~$a_0$ and $\phi = 0^{\circ}$, in addition to the vicinity of the 3 local minima, one can see larger energy barriers as well ($V>100$~cm$^{-1}$) at other $\theta$-orientations, referring  to large angular anisotropies.

\begin{figure*}
\includegraphics[width=0.50\linewidth]{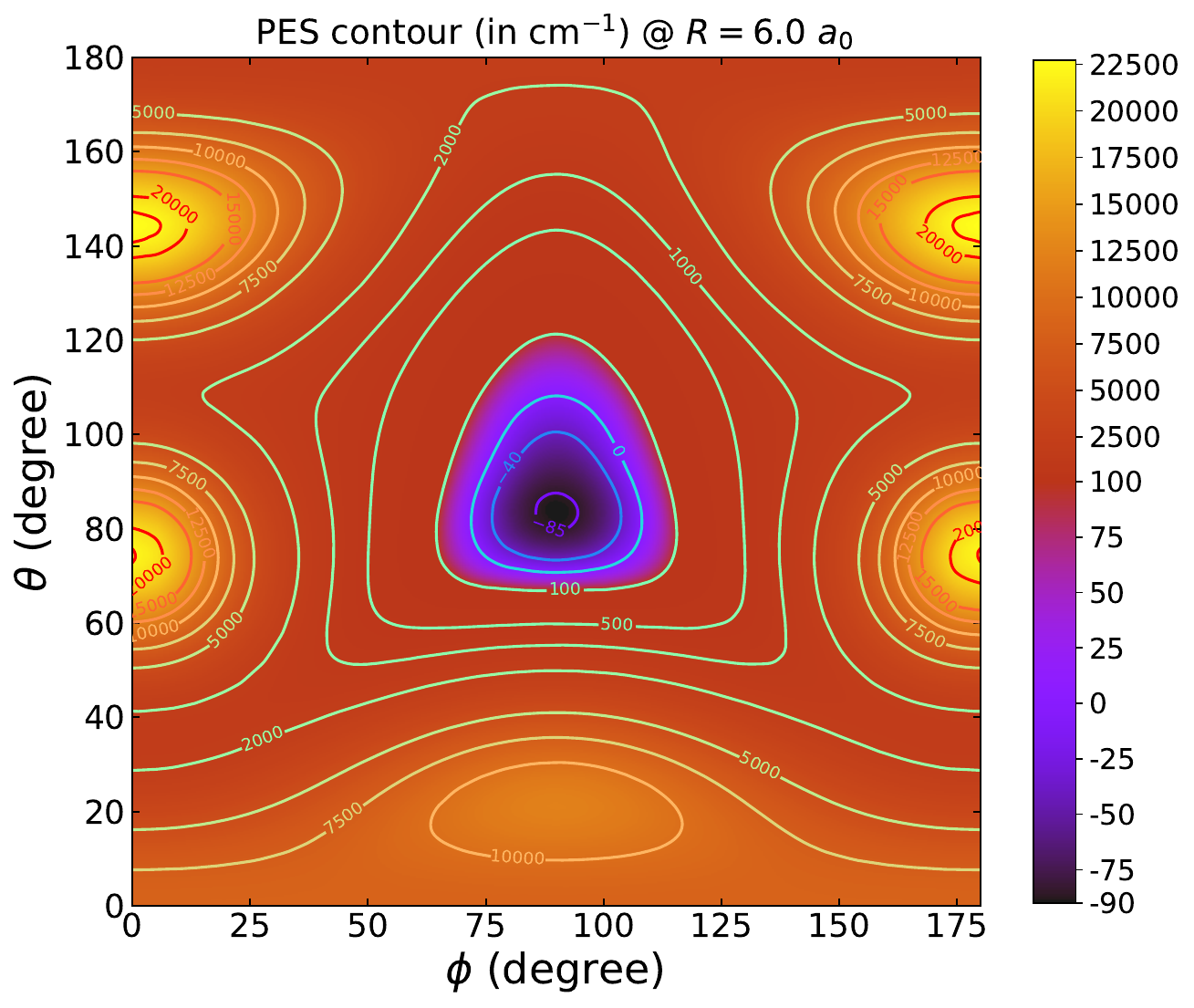}
\includegraphics[width=0.482\linewidth]{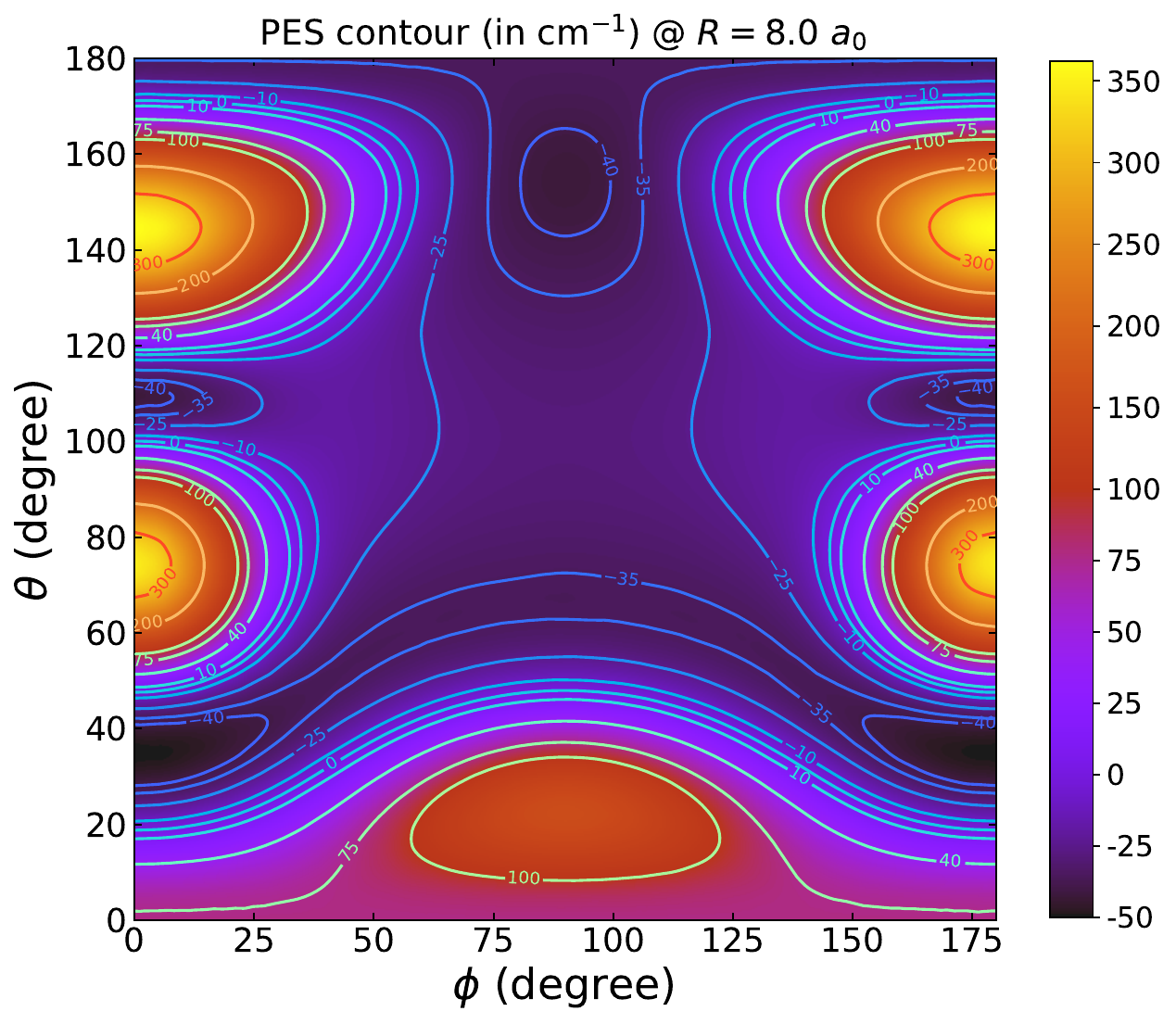}
\caption{Contour plots representing the $\theta$ $vs.$ $\phi$ dependence of the $c$-C$_5$H$_6 + $ He PES at fixed radial distances $R=6.0$~$a_0$ (left) and $R=8.0$~$a_0$ (right).}
\label{fig:PEScontPT}
\end{figure*}

The contour plots in Fig.~\ref{fig:PEScontPT} help to understand more the general angular anisotropy of the PES. They show the $\theta$ $vs.$ $\phi$ behaviour at some fixed radial distances ($R =6.0$~$a_0$ on the left subplot and $R =8.0$~$a_0$ on the right). As one can see on the former, the PES is dominantly repulsive at short distances, except for a narrow region at intermediate angles in the surrounding of the global minimum. By going out from the well, the repulsive force is increasing very intensively, and we can reach some high-energy centers (with $V > 20000$~cm$^{-1}$), which refer to the vicinity of hydrogen sites of C$_5$H$_6$. Increasing the radial distance by 2~$a_0$ leads to a really different surface (see the right subplot of Fig.~\ref{fig:PEScontPT}). First, it is notable that we do not observe any highly repulsive regions, since the He projectile will be sufficiently far from any atomic centers of the target. However, we still see a clear picture about the location of the H-cores of C$_5$H$_6$, where the repulsive regions with  $V > 350$~cm$^{-1}$ draw the contour of these atomic sites arranged in a highly-symmetric pentagon shape, which corresponds to the backbone of cyclopentadiene. The attractive part of the PES is very broad here, filling the major part of the angular space. We found again several local minima where the interaction potential is below $-40$~cm$^{-1}$. At intermediate angles, the PES contours highlight a wide attractive region, but it is not centered around the global minimum any more. Both subplots exhibit a perfect left-right symmetry with respect to $\phi = 90^{\circ}$ due to the $C_{2v}$-symmetry of the molecule. As one can see in both Figs.~\ref{fig:PEScontRT} and \ref{fig:PEScontPT}, the iso-contour lines exhibit a smooth behavior with respect to all coordinates, which is also a good feedback on the quality of the analytical fit. Due to the pronounced well and the multiple secondary minima along the PES, quantum-dynamical calculations require a fine resolution over collision energies because of the heavy resonances related to the formation of numerous bound and quasi-bound states.

\section{\label{sec:ScattMeth} Scattering calculations}

Once the analytical PES has been successfully implemented in the \texttt{MOLSCAT} scattering code,\cite{Hutson_2019} we have calculated state-to-state cross sections for the rotational (de-)excitation of $c$-C$_5$H$_6$ by He impact. The total energies have been varied from the lowest inelastic threshold up to $60$~cm$^{-1}$. We used the exact close-coupling (CC) quantum scattering method for the calculations.\cite{Arthurs_Dalgarno1960} It is worth emphasizing that, to the best of our knowledge, the current work is the first study where the CC method has been systematically applied for an atom + molecule collision problem involving more than ten atoms (the collisional complex in this work contains 12 atoms). For this reason, we also perform calculations using the approximate coupled states (CS) scattering method, which has been more often used to treat the collisional excitation of heavy and complex species. The CS method is based on an assumption that the coordinate system can be rotated so that the basis functions associated to the projection of the rotor momentum to the main rotational axis, $j_z$, can be decoupled (it was originally called $j_z$-conserving coupled states approximation).\cite{McGuire_1974} In other words, it neglects the Coriolis couplings, and due to this, the number of coupled equations (channels) can be significantly reduced.

The rotational structure of cyclopentadiene has been discussed in Section~\ref{sec:Intro} in details. In the dynamical studies we used the experimental rotational constants,\cite{bogey1988} thus $A = 0.281$ cm$^{-1}$, $B = 0.274$ cm$^{-1}$ and $C = 0.142$ cm$^{-1}$. We targeted all {\it ortho-} and {\it para-}C$_5$H$_6$ nuclear spin states with internal energies up to $\sim 10$~cm$^{-1}$, including a total of 22 rotational levels per symmetry and all states with $j \leq 5$, along with some of those with $j = 6,7$. The reduced mass of the [C$_5$H$_6 -$He] complex is equal to 3.774076 atomic mass unit.

The efficiency of the scattering calculations has been increased by finding an optimal truncation on the rotational basis set ($j_\mathrm{max}$) and largest total angular momentum ($J_\mathrm{tot}$). These were selected based on systematic convergence test calculations depending on the collision energy. The convergence threshold criteria was set to a maximum of $1.0\%$ mean absolute error for $j_\mathrm{max}$ and $0.005\%$ for $J_\mathrm{tot}$. Obeying such convergence criteria, the highest rotational levels which were included in our calculation are those with $j_\mathrm{max} = 20$ and the largest angular momenta reached $J_\mathrm{tot} = 36$. The parameters of the numerical propagation, such as the initial ($R_\mathrm{min}$) and final radial distance ($R_\mathrm{max}$) along with the switching point between the log-derivative and Airy propagators ($R_\mathrm{mid}$), have also been optimized with a convergence criteria $<0.3 \%$. A cutoff parameter ($E_\mathrm{max}$) was used to limit the contribution from non-significant high-lying rotational levels, which was also parametrized through systematic tests with a convergence criteria $<0.3 \%$. To properly resolve the resonances in the cross sections, a very small energy step size has been applied: $0.01$~cm$^{-1}$ at sub-wavenumber kinetic energies, $0.1$~cm$^{-1}$ up to $50.0$~cm$^{-1}$ total energy, and $0.2$~cm$^{-1}$ at the highest energies.

Based on the summarized errors from the {\it ab initio} calculations, the analytical fit and the optimizations in the scattering calculations, our estimations show that the accuracy of the collisional data presented in this work is always better than $\sim 10\%$, and is generally on the level of $1-2\%$ for all particular energies and rotational transitions considered.

\section{\label{sec:ResDis} Discussion}

\begin{figure*}[ht]
\centering
\includegraphics[width=0.49\linewidth]{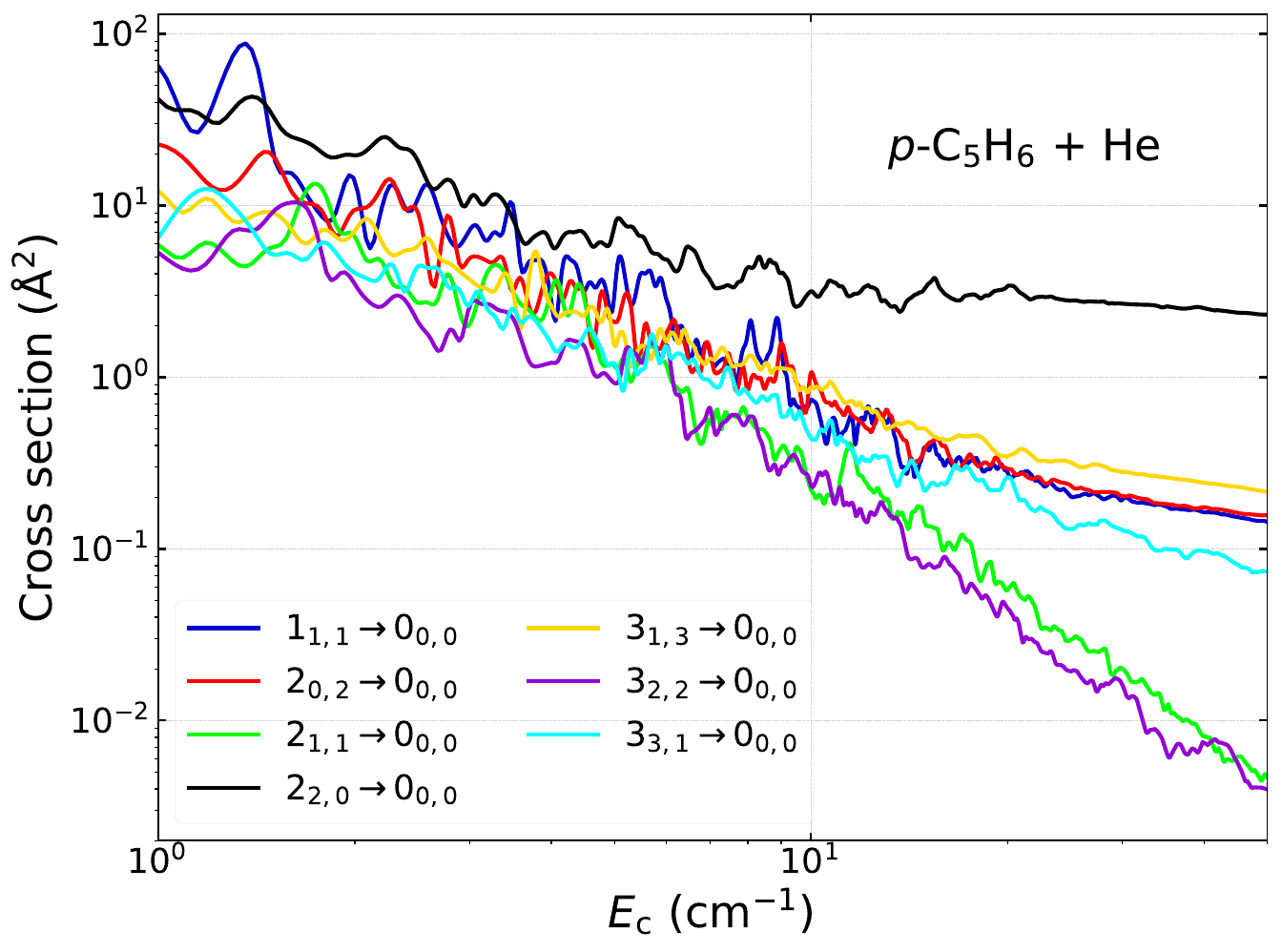}
\includegraphics[width=0.49\linewidth]{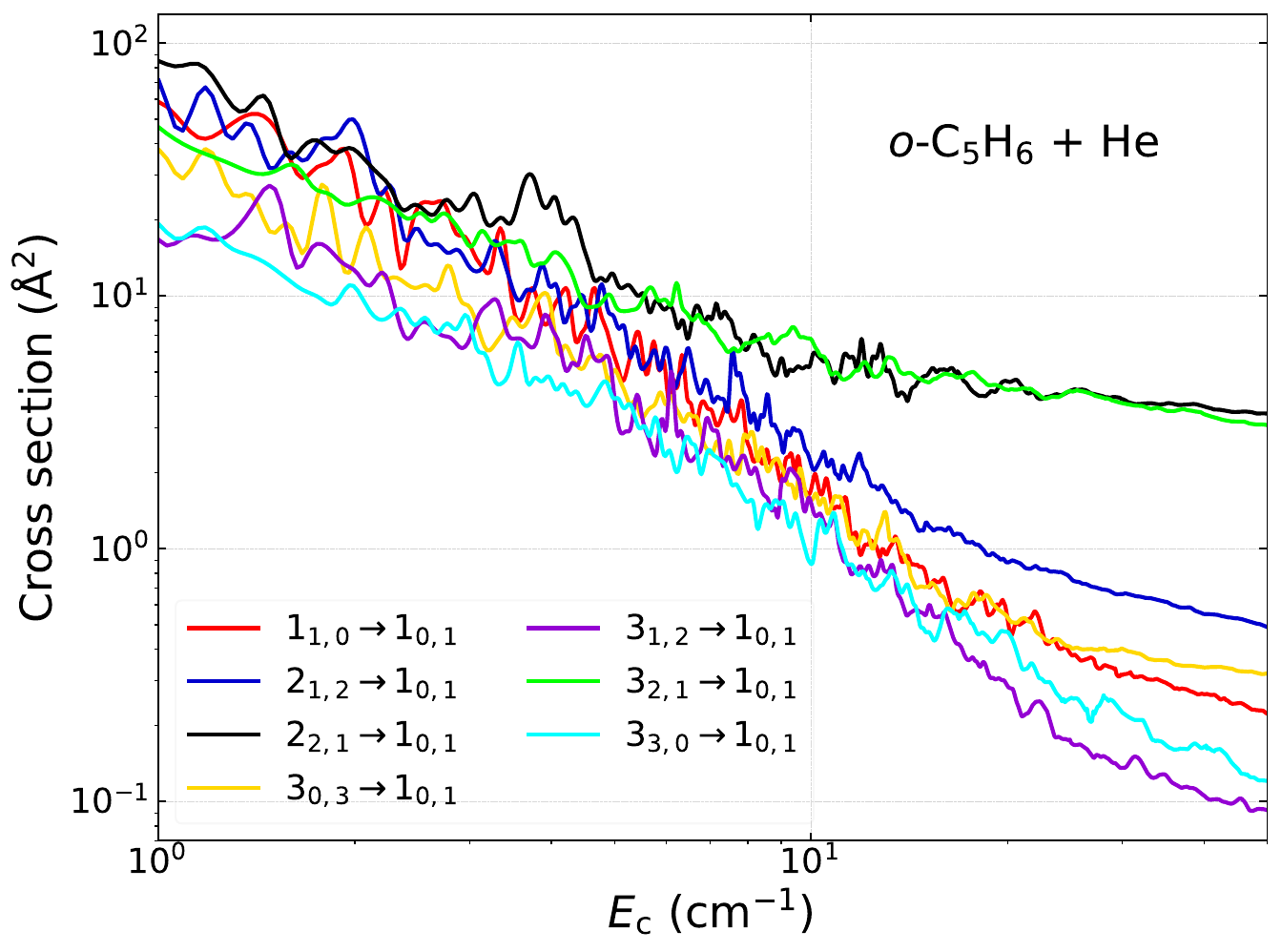}
\caption{Collision energy dependence of the rotational de-excitation cross sections for transitions towards the lowest {\it para}-C$_5$H$_6$ ($0_{0,0}$, left) and {\it ortho}-C$_5$H$_6$ ($1_{0,1}$, right) states due to collisions with He.}
\label{fig:XS_low}
\end{figure*}

\begin{figure*}
\centering
\includegraphics[width=0.49\linewidth]{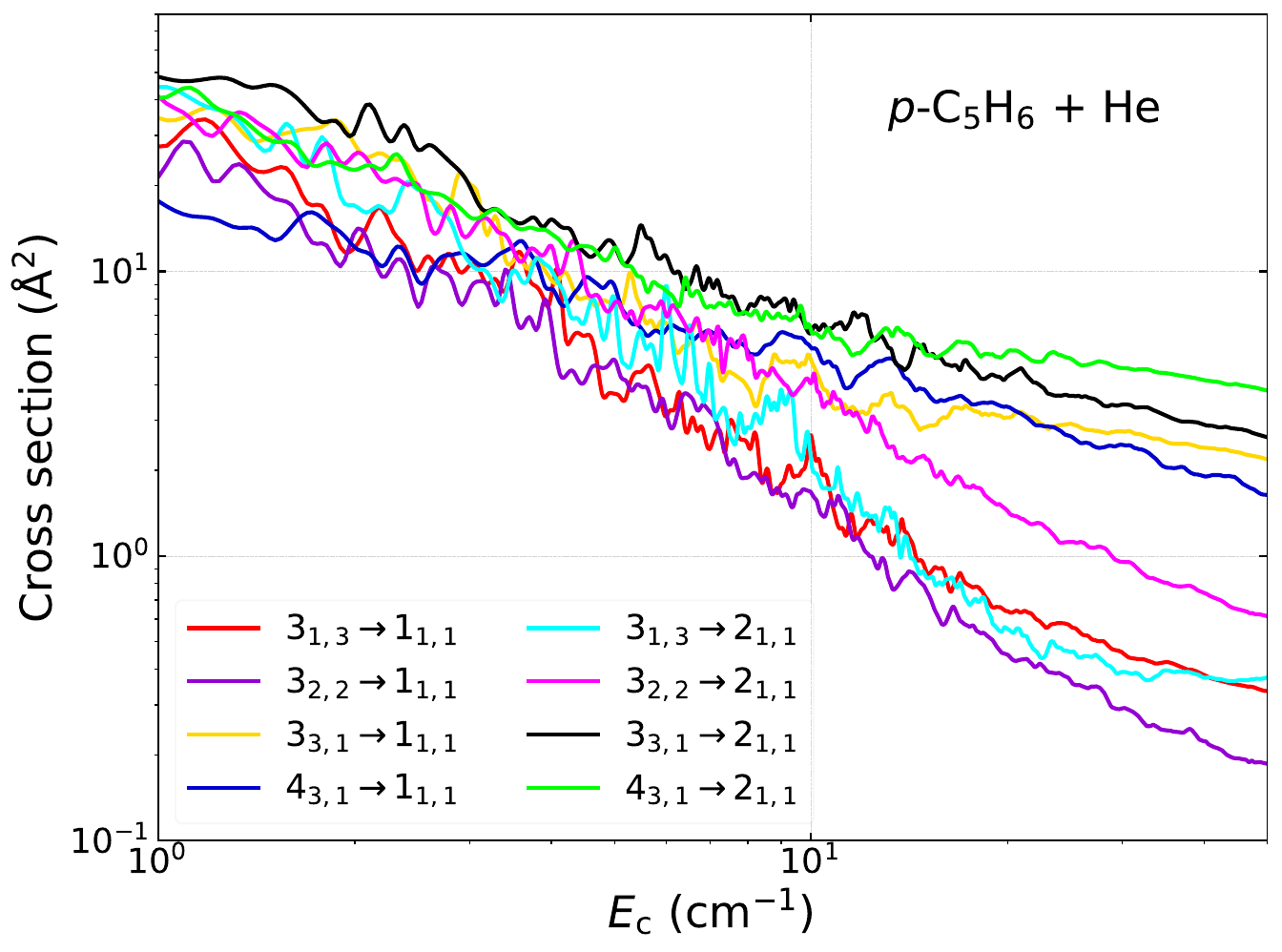}
\includegraphics[width=0.48\linewidth]{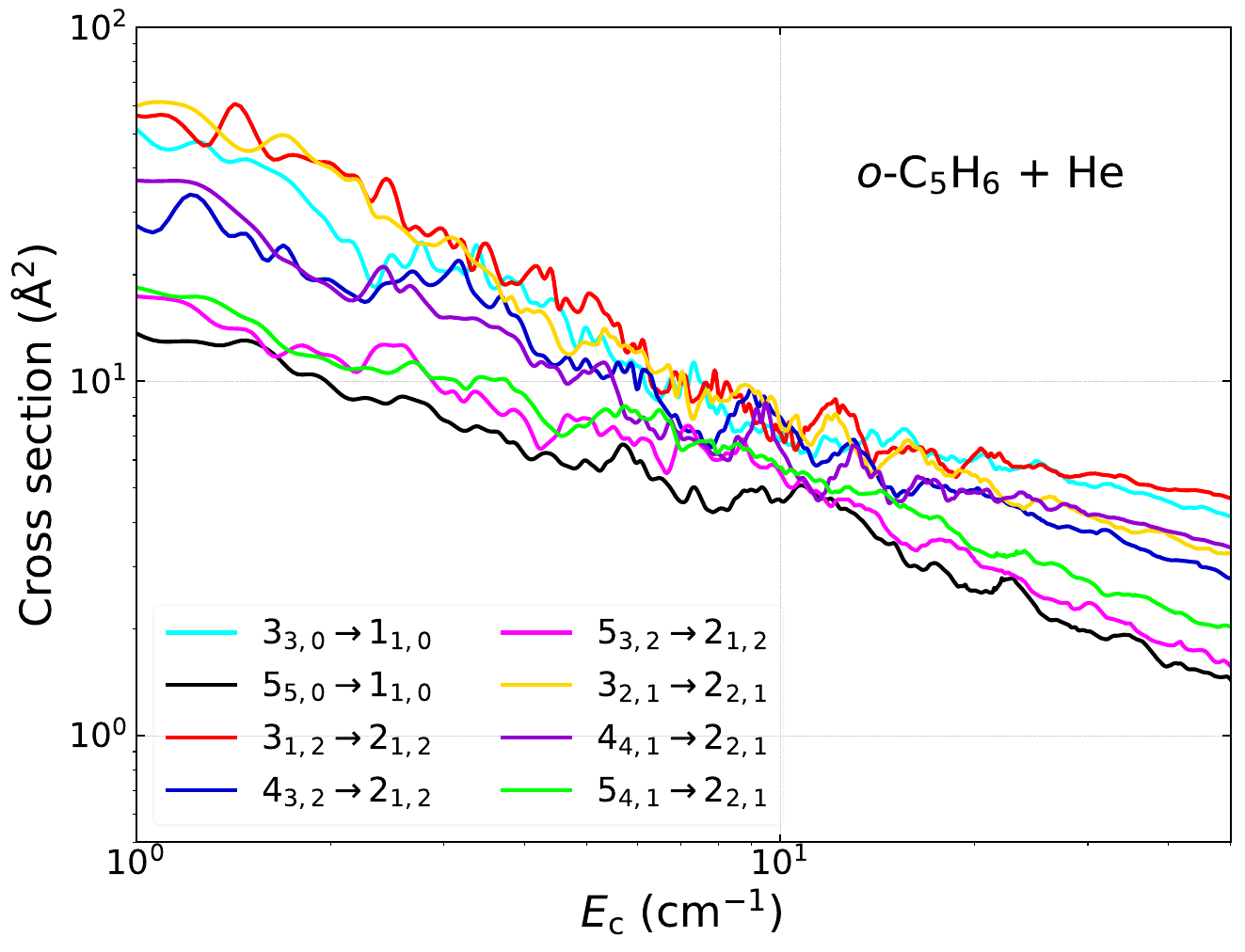}
\caption{Same as Fig.~\ref{fig:XS_low} but for various other transitions in collisions of {\it para}-C$_5$H$_6$ (left) and {\it ortho}-C$_5$H$_6$ (right) with He.} %(00,0) from all J = 5Ka,Kc (left) and J = 6Ka,Kc (right) levels (para−C5H6 + He collision).
\label{fig:XS_prop}
\end{figure*}

We have analyzed the energy dependence and the propensity rules in the (de-)excitation cross sections $\sigma(j_{k_a,k_c} \rightarrow j'_{k'_a,k'_c})$ for the $c$-C$_5$H$_6 + $ He collision in Figs.~\ref{fig:XS_low}-\ref{fig:XS_prop}. First, as it is expected due to the small splitting between the rotational levels, all cross sections exhibit a dense resonance structure. These are a combination of both Feshbach-type and shape (orbiting) resonances, which are typical for such loosely bound complexes. As the collision energy increases, the resonances become less pronounced.

Fig.~\ref{fig:XS_low} shows the variation of the cross sections for transitions towards the lowest rotational levels, thus $j_{k_a,k_c} = 0_{0,0}$ in the case of {\it para}-C$_5$H$_6$ and $j_{k_a,k_c} = 1_{0,1}$ in the case of the {\it ortho} nuclear spin symmetry. As one can see, it is not straightforward to identify general propensity rules with respect to the rotational quantum numbers. For example, we do not observe any significant trends in the case of $\Delta j = 0, \pm1$  transitions, which are often favoured in the case of other nonlinear species (for example, see Ref.\cite{Demes_2023}). This is also the case when we examine the propensities with respect to the $k_a$ projection quantum number, since neither $\Delta k_a = 0 $ nor $\Delta k_a = 1 $ transitions are favoured. It is worth highlighting however that we can identify a strong and systematic propensity trend in the case of $k_c$-conserving transitions ({\it i.e.} when $\Delta k_c = 0$). Above $\sim 10$ cm$^{-1}$ collision energies, most of the cross sections tend to monotonically decrease, except for those that correspond to this rule. In the case of {\it para}-C$_5$H$_6$ (Fig.~\ref{fig:XS_low} left subplot), such $k_c$-conserving transition is the one from the $2_{2,0}$ to the $0_{0,0}$ rotational level, while for {\it ortho}-C$_5$H$_6$ (right subplot), the corresponding ones are the $2_{2,1} \rightarrow 1_{0,1}$ and $3_{2,1} \rightarrow 1_{0,1}$ transitions. As one can see, the slope of their energy dependence is not too steep, keeping their magnitude high even when other transitions are strongly decreasing. Due to this, such favoured transitions are usually by more than an order of magnitude larger compared to the less dominant ones, which will be obviously visible in the collisional rate coefficients too, especially at higher kinetic temperatures.

In Fig.~\ref{fig:XS_prop} we illustrate the collisional propensities in the case of transitions between various initial and final states, also including some levels with $j=4$ and $j=5$. The proper analysis of the propensity rules are important to understand the relative branching between the (de-)excitation channels and to estimate the relative strengths of the transitions based on the corresponding quantum numbers even in the case of high-lying rotational states.\cite{Demes_2023} As one can see, the overall variation of the cross sections with collision energy shows similar decreasing trends, but the most dominant transitions usually have a smaller slope. First, we examine the transitions in {\it para}-C$_5$H$_6$ towards the $1_{1,1}$ and $2_{1,1}$ final states in the left subplot of the figure. We provide transitions from all states with $j=3$ and additionally those from $4_{3,1}$. As one can see again, there are no systematic rules to favour any of the $\Delta j$ or $\Delta k_a$ transitions, however there is some slight dominance of the $\Delta j =1,2$ transitions over the one with $\Delta j = 3$. This is most probably related to the dominant $V_{2m}$ expansion terms of the PES, which have the most significant contribution in the couplings which drive the $\Delta j =1,2$ processes (with the exception of the $j - j' = 1 - 0$ transition, where only $V_{1m}$ contributes directly). We observe the most pronounced propensities in the case of the $\Delta k_c = 0$ processes again, {\it i.e.} the transitions from the $3_{3,1}$ and $4_{3,1}$ are the strongest. This effect is more visible at energies above $\sim 10$ cm$^{-1}$, leading to differences up to about an order of magnitude at the highest energies between the less and more dominant processes. The similar trends between the two initial states with $k_c = 1$ confirms that it is not related to the connection between the other quantum numbers, since it is observed both when $j = k_a$ and $j \neq k_a$.

In the right subplot of Fig.~\ref{fig:XS_prop} we systematically examine the cross sections for {\it ortho}-C$_5$H$_6$ with $k_c$-conserving transitions. Their energy dependence is rather similar in general, where their magnitude is usually high and does not drop strongly as the energy increasing. For example, if we compare them at $50$~cm$^{-1}$, we see that their magnitudes are always large, between $1$  and $10$~\AA{}$^2$, while all other transitions tend to be closer to $0.1$~\AA{}$^2$ or sometimes even below $0.01$~\AA{}$^2$ at this collision energy. We observe some slight, but systematic differences however, which demonstrate again the favour of $\Delta j =1,2$ processes over those with $\Delta j =3,4$ (for example, compare the transitions from the $4_{3,2}$ and $5_{3,2}$ states to the $2_{1,2}$ final state). It is important to point out that really significant gaps between the cross sections are observed only above $\sim 10$ cm$^{-1}$. We can also see that these are not correlated to the energy gap between the initial and final states, but more likely driven by the collisional propensity rules, dominantly with respect to $k_c$.

To understand the origin of the observed propensities, we examined their presence in the case of other molecular collisions too. A good candidate for this can be the cyclic C$_3$H$_2$, which, analogously to $c$-C$_5$H$_6$, also has a carbon-cycle backbone and a $C_{2v}$ rotational symmetry with a similar partitioning to {\it ortho}/{\it para} nuclear spin species (note that both benzonitrile and propylene oxide have very different rotational structure, despite their asymmetric top rotational symmetry). We carried out thus an analysis based on the work of Ben Khalifa {\it et al.} (2019)\cite{Khalifa2019} devoted the rotational excitation of C$_3$H$_2$ in collision with He. We performed a state-to-state comparison of cross sections involving various initial and final levels (with $j \leq 5$), but we did not find any propensity trends with respect to $k_c$. To complete our analysis, we also examined these trends in the case of other, smaller asymmetric top species, which have been studied more extensively. We analyzed the rotational excitation data reported for H$_2$O,\cite{Faure_2007} NH$_2$,\cite{Bouhafs_2017} H$_2$S,\cite{Dagdigian_2020a} and HCO.\cite{Dagdigian_2020b} These works reveal however very different results: while for H$_2$O and NH$_2$, some propensities are found in terms of all quantum numbers $(\Delta j, \Delta k_a, \Delta k_c = 0,\pm1)$, in the case of H$_2$S, only the $\Delta j = 0$ transitions are really distinct from others, and in the case of HCO, the strongest propensity rules are found for $\Delta k_a = 0$ transitions. Our analysis confirms that the trend towards the conservation of the $k_c$ projection in rotational excitation is not a universal tendency for $C_{2v}$ asymmetric top species, but it is rather a feature relevant for cyclopentadiene. The origin of such propensity however requires more detailed investigations.

\begin{figure}
\centering
\includegraphics[width=0.7\linewidth]{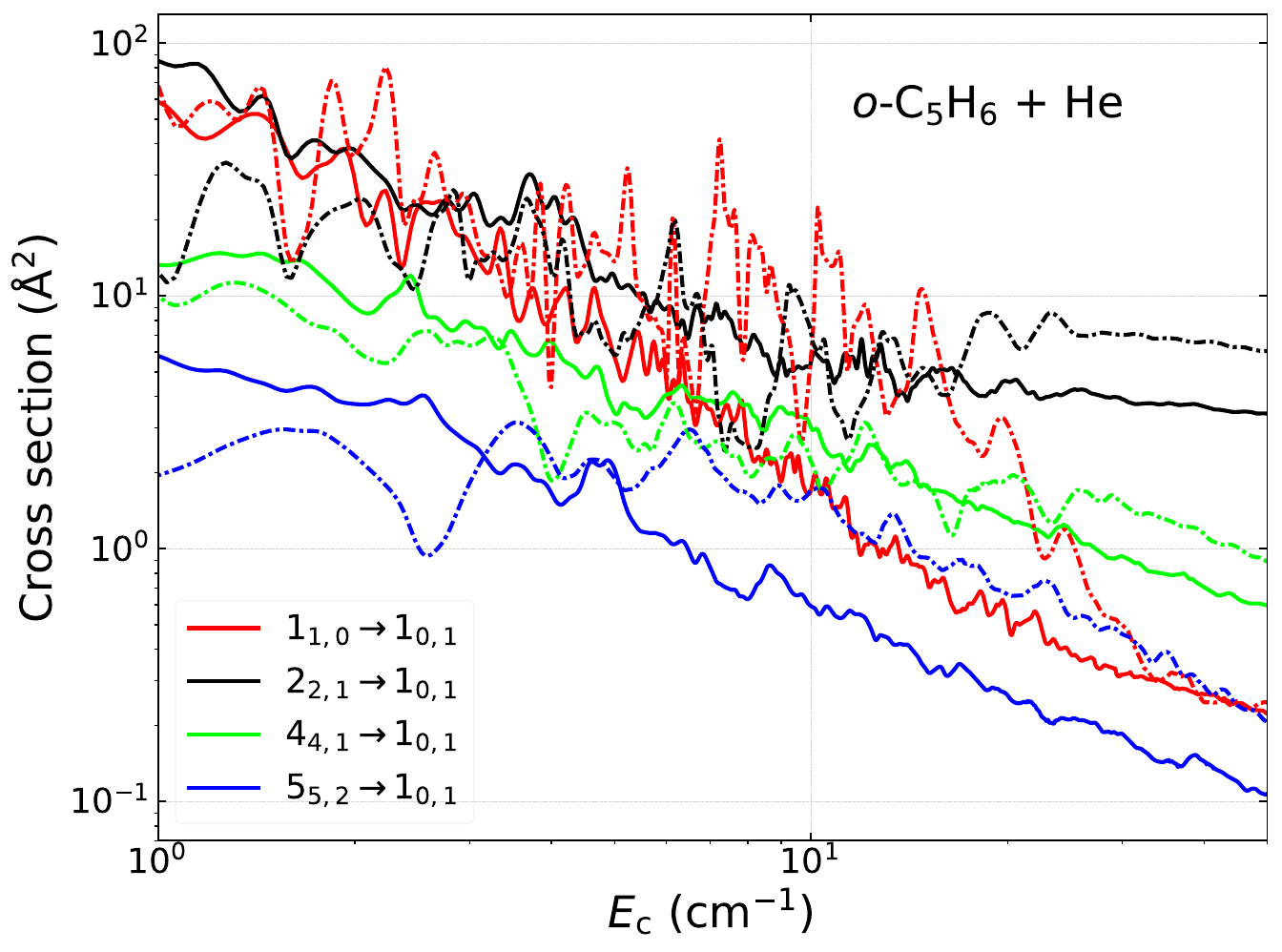}
\caption{Variation of the cross sections calculated by the close coupling, CC (solid lines) and the coupled states, CS (dash-dot lines) scattering theories for some rotational transitions towards the $1_{0,1}$ state of {\it ortho}-C$_5$H$_6$ due to collision with He.}
\label{fig:XS_CCvsCS}
\end{figure}

Since this is the first time the rotational excitation of a large cyclic species is studied by means of the CC method, we would like to estimate its accuracy in contrast with the CS approximation, which was used previously for many complex species, including propylene oxide\cite{Faure_2019} and benzonitrile.\cite{Khalifa_2023} For this purpose, we also preformed systematic cross section calculations by the coupled states approximation using the same parameters as in the case of the CC theory. Fig.~\ref{fig:XS_CCvsCS} shows the comparison for some randomly selected transitions towards the $1_{0,1}$ rotational state of {\it ortho}-C$_5$H$_6$, where solid lines represent the results from the CC calculations, while the dash-dot lines are for those from the CS approximation. Notable differences can be seen between these cross sections, both in their magnitude and in their variation with collisional energy. The CC data strongly deviate from the CS ones in the whole energy interval, apart from some low-energy regions (from $\sim 3$ up to $10-20$ cm$^{-1}$, depending on the transition), where they are usually comparable. A very important finding is that at higher collision energies, which are the most crucial when deriving rate coefficients for higher temperatures, the two scattering theories provide significantly different cross sections. The observed differences are ranging from a few percents up to about a factor of two, which was also confirmed in the case of other de-excitation processes not shown in the figure. It is worth to emphasise that the deviations are also significant (within a factor of $1.5-2$) for the most dominant $k_c$-conserving transitions, such as the $2_{2,1} \rightarrow 1_{0,1}$ or $4_{4,1} \rightarrow 1_{0,1}$ in Fig.~\ref{fig:XS_CCvsCS}. In some cases, the resonance behaviour is wrongly described in the cross sections from the CS approximation, leading to unphysically large structures, which may lead to additional uncertainty factors, as it can be seen in the case of the fundamental $1_{1,0} \rightarrow 1_{0,1}$ transition for example. For other processes, a quasi-constant gap ({\it i.e.} a constant scaling factor) tends to be present above certain energies between the results calculated from the two theories. Whether this scaling applies or not at higher collision energies, should be further investigating however, extending both the CC and CS calculations towards higher energies, which is extremely costly in the case of the close coupling method.

\section{\label{sec:Concl} Conclusions}

We have presented state-to-state rotational (de-)excitation cross sections for the collision of {\it c}-C$_5$H$_6$ with He atom. To the best of our knowledge, this is the first time that the collisional excitation of a large cyclic species with more than 10 atoms is systematically studied using the exact close-coupling formalism. In order to perform scattering dynamics calculations, we have developed a 3D rigid-rotor potential energy surface for the collisional system, calculated from the explicitly correlated CCSD(T)-F12b {\it ab initio} theory with an aug-cc-pVTZ atomic basis set. The global minimum of the [$c$-C$_5$H$_6 -$He] complex is about $-90$~cm$^{-1}$ and located around $R \simeq 6.0$~$a_0$.

Using this analytical PES, we calculated the (de-)excitation cross sections between the lowest 22 states of both the {\it ortho}- and {\it para}-C$_5$H$_6$ nuclear spin species, {\it i.e.} those with internal energies below $\sim 10$~cm$^{-1}$ (including all levels with $j=5$ and some of those with $j=6,7$). We studied a range of collision energies up to $50$~cm$^{-1}$, where a series of intensive Feshbach- and shape-resonances have been found for all particular transitions. The collisional propensity rules have also been analysed comparing the cross sections with respect to the variation of the rotational quantum numbers between the initial and final states. The analysis showed a systematic favour of the $k_c$-conserving transitions, while there are no general trends found for any particular $\Delta j$ or $\Delta k_a$ processes. These rules apply for both nuclear symmetries equally.

It is worth highlighting that we found significant differences of up to a factor of 2 in contrast with the results calculated by the approximate coupled state scattering theory, which is generally used to treat heavy collisional systems. The deviations are also large at higher collision energies as well as in the case of the most dominant $\Delta k_c = 0$ transitions. These findings suggest that the collisional rate coefficients calculated form the CS model can have larger uncertainties at low temperatures, below $10-15$~K, which will have a notable effect on the astronomical modelling of cold interstellar clouds, where the typical temperature is $<10$~K. To undoubtedly verify this conclusion however, further studies are needed, with a significant extension of the range of collision energies and the number of rotational levels, followed by proper radiative transfer modelling of the C$_5$H$_6$ emission lines under non-LTE conditions. According to our estimations, the accuracy of the cross sections calculated in the present work is always better than $\sim 10 \%$ and usually as low as a few percents.

\section*{Supplementary Material}
The expansion coefficients of the C$_5$H$_6 +$He potential energy surface are available within the electronic supplementary material (available at \url{https://www.rsc.org/suppdata/d4/cp/d4cp01380h/d4cp01380h1.zip})

\section*{Conflicts of Interest}
There are no conflicts to declare.

\section*{Acknowledgements}
We acknowledge financial support from the European Research Council (Consolidator Grant COLLEXISM, Grant Agreement No. 811363) and the Programme National ‘Physique et Chimie du Milieu Interstellaire’ (PCMI) of CNRS/INSU with INC/INP cofunded by CEA and CNES. We wish to acknowledge the support from the CEA/GENCI for awarding us access to the TGCC/IRENE supercomputer within the A0110413001 project. This article is based upon collaborations supported by the COST Actions CA18212–Molecular Dynamics in the GAS phase (MD-GAS) and CA21101–Confined Molecular Systems: From a New Generation of Materials to the Stars (COSY), supported by COST (European Cooperation in Science and Technology). F.L. acknowledges the Institut Universitaire de France. The authors are very grateful to J. Cernicharo, M. Ag\'{u}ndez, J.~Loreau and G.~Czak\'{o} for the essential and fruitful discussions.

%%%END OF MAIN TEXT%%%

%If notes are included in your references you can change the title from 'References' to 'Notes and references' using the following command:
%\renewcommand\refname{Notes and references}

%%%REFERENCES%%%
\bibliography{references} %You need to replace "rsc" on this line with the name of your .bib file
\bibliographystyle{rsc} %the RSC's .bst file

\end{document}